\documentclass[aps,twocolumn,pra,superscriptaddress,amsmath,showpacs,tightenlines]{revtex4-1}
\usepackage{amssymb}
\usepackage{amsmath}
\usepackage{graphicx}
\usepackage{subfigure}
\usepackage{epsfig}
\usepackage{amsfonts}
\usepackage{mathrsfs}
\usepackage{xcolor}
\usepackage[toc,page,title,titletoc,header]{appendix}
\begin{document}

\title{High-fidelity Generation of Bell and W States in Giant Atom System via Bound State in the Continuum}

\author{Mingzhu Weng}
\affiliation{Center for Quantum Sciences and School of Physics, Northeast Normal University, Changchun 130024, China}
\author{Hongwei Yu}
\affiliation{Center for Quantum Sciences and School of Physics, Northeast Normal University, Changchun 130024, China}
\author{Zhihai Wang}
\email{wangzh761@nenu.edu.cn}
\affiliation{Center for Quantum Sciences and School of Physics, Northeast Normal University, Changchun 130024, China}

\begin{abstract}
In this paper, we propose a high-fidelity scheme for generating entangled states in a system of two and three giant atoms coupled to the coupled resonator waveguide. Our approach leverages the bound state in the continuum, which is robust against waveguide disorder. Specifically, we achieve a fidelity exceeding $98\%$ for Bell state generation, overcoming the limitations of conventional decoherence-free interaction mechanisms. This scheme can be readily extended to a three-giant-atom system for generating W states. In both the two- and three-atom setups, the maximally entangled states are generated in a short time and remain stable even as time approaches infinity. Our proposal is feasible for implementation on state-of-the-art solid-state quantum platforms and significantly broadens the applications of giant atoms and waveguide QED system in quantum information processing.
\end{abstract}


\maketitle
\section{introduction}
Quantum entanglement lies at the core of the quantum information and plays a significant role in quantum communication and quantum networks~\cite{RH2009,LA2015}, such as quantum key distribution~\cite{AK1991,FG2003}, quantum teleportation~\cite{CH1993,JJ2003}, and quantum-computing technology~\cite{DP2000}. Among the various types of entangled states, Bell states represent the maximally entangled bipartite states~\cite{PG1995,LQ1999}, while W states are a prominent class of many-qubit entangled states known for their exceptional resilience to particle loss~\cite{WD2000,MK2000}. To date, the Bell state and W state have been extensively studied~\cite{HY2020,MA2009,MN2010,JP2021} and their suitability for applications in quantum information protocols has driven their preparation in various systems~\cite{FH2014,AA2013,CL2015,BF2019}.

In the early days, the preparation of entangled states primarily focused on photonic and ionic systems~\cite{RP2009,WW2009,HH2005}. The Rydberg ensemble has emerged as a promising alternative for achieving scalable entanglement among neutral atoms~\cite{VM2021,AO2019,YJ2020,MG2015,JLi2019,CW2022}. Currently, the advancement of quantum information in solid-state systems has motivated researchers to explore effective schemes for realizing entangled states in artificial systems, such as semiconductor quantum dots \cite{WB2012} and superconducting qubits \cite{JQ2011}. Especially, superconducting circuits are regarded as one of the most promising solid-state systems for implementing quantum computing. In contrast to natural atoms, the superconducting qubits can be engineered to couple with the waveguide via multiple points to acts as giant artificial atoms~\cite{MV2014,RM2017,AM2021}. The interference effects arising from these multiple coupling points \cite{BK2020,YT2022,XL2022,AF2018,LD2023,DC2020,AC2020,AS2022} can lead to the emergence of bound states in the continuum (BIC)~\cite{LL2024,ER2024}. The so called BIC locates in the continual band of the system in the frequency domain, but the wave function exhibits localized character in the real space. Nowadays, the BIC is widely applied in laser technology~\cite{KI2019,KK2017} and quantum information processing \cite{XZ2023}.

In this paper, we propose a scheme for generating maximally entangled states in a giant atom system that is coupled to a coupled resonator waveguide~\cite{BS2018,MZ2019,JL2019,JS2020,SP2021,KF2013}. We demonstrate that the BIC in our system exhibits robustness against both onsite frequency and hopping strength disorder of the waveguide. This robustness provides a distinctive mechanism for entangled state generation. Actually, in giant atom system, the maximum entangled state can also be prepared by leveraging the adjustable decay processes and interference processes of giant atoms in waveguides~\cite{XL2023,AC2023}. Alternatively, the Bell and W states generation via steady state is also an effective approach, but it requires sophisticated designs for the coupling and dissipation terms~\cite{GZ2023,GZa2023}. In contrast, in our scheme based on the BIC, we only need to find the optimal driving duration to achieve the maximum entangled state, which can be maintained after undergoes a long time evolution.

For the two-giant atom setup, we demonstrate that the Bell state can be realized with a fidelity higher than $98\%$ in braided, separate and nested configurations. These findings
challenge the notion that only the decoherence free interaction (DFI)~\cite{AF2018,LD2023}, which is exclusively engineered in the braided setup, is useful in quantum information processing.
Furthermore, we extend our analysis to a configuration involving three giant atoms, to obtain the high-fidelity W states with the assistance of BIC.

\section{Model}

\begin{figure}
\centering
\includegraphics[width=1\columnwidth]{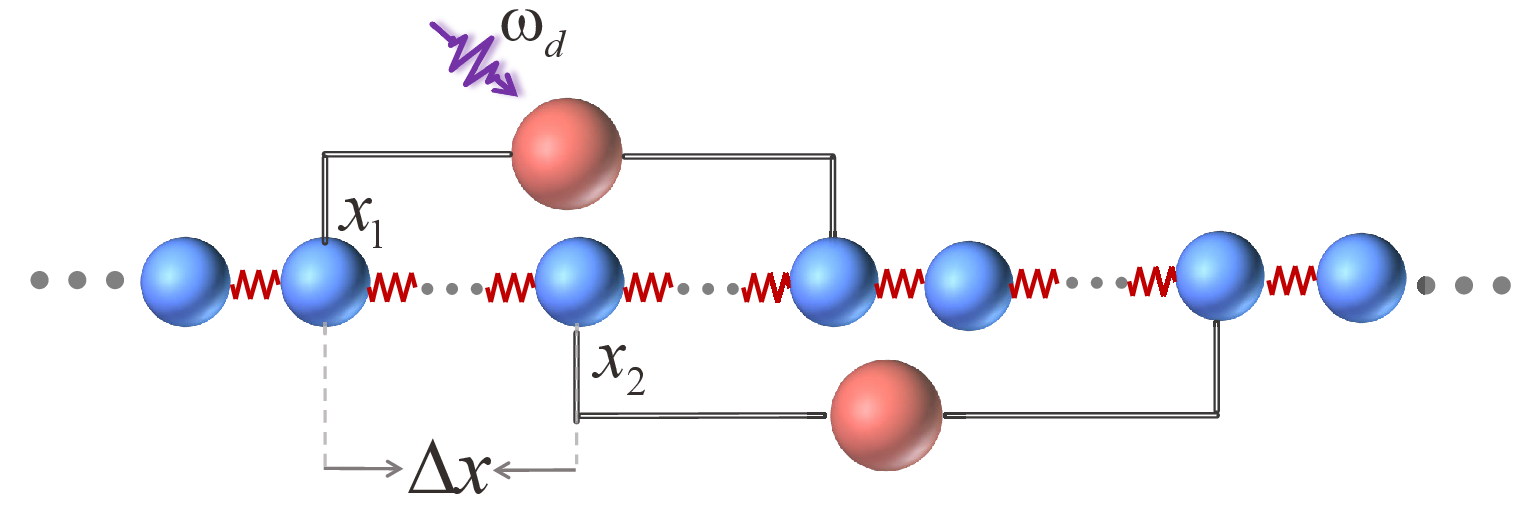}
\caption{Schematic configuration for two giant atoms which couple to a one-dimensional coupled-resonator waveguide via two sites in a braided configuration. The blue balls depict the resonator in the waveguide and the red balls are the giant atom. The left coupling site of the $i$th giant atom is $x_{i}$ $(i=1,2)$.}
\label{twodevice}
\end{figure}

The model in our consideration is composed by two giant atoms which couple to a coupled resonator waveguide in braided configuration, as illustrated in Fig.~\ref{twodevice}.  The giant atoms are designed to couple with the waveguide through two separate coupling sites $x_{i}$ and $x_{i}+n$ $(i=1,2)$. The coupling space between the left leg of two giant atoms is $\Delta x=x_{2}-x_{1}$ in Fig.~\ref{twodevice}. The total Hamiltonian, with the rotating wave approximation, contains three parts of $H=H_a+H_c+H_{I}$ ($\hbar=1$), where
\begin{equation}
H_{a}=\Omega\sum_{i=1,2}\sigma_{i}^{+}\sigma_{i}^{-}
\end{equation}
defines the free Hamiltonian of the two giant atoms. $\Omega$ is the transition frequency between the excited state $|e\rangle_{i}$ and the ground state $|g\rangle_{i}$ of the giant atom $i$. The operator $\sigma_{i}^{+}(\sigma_{i}^{-})$ is the raising (lowering) operator of the giant atom. The coupled resonator waveguide is modeled by the Hamiltonian
\begin{equation}
H_c=\omega_{c}\sum_{j}a_{j}^{\dagger}a_{j}-\xi\sum_{j}(a_{j+1}^{\dagger}a_{j}+a_{j}^{\dagger}a_{j+1}).
\label{Hwaveguide}
\end{equation}
Here, $a_{j}^{\dagger}(a_{j})$ is the creation (annihilation) operator of the field in the waveguide on site $j$. $\omega_c$ is the resonators' frequency and $\xi$ is the nearest-neighbor hopping strength. Finally, the coupling between the giant atoms and the waveguide field is described by the Hamiltonian
\begin{equation}
H_I=\sum_{i=1,2}g[\sigma_{i}^{+}(a_{x_{i}}+a_{x_{i}+n})+\rm{H.c.}]
\end{equation}
where $g$ is the coupling strength.

\label{Twoatom}

\begin{figure}
\centering
\includegraphics[width=1\columnwidth]{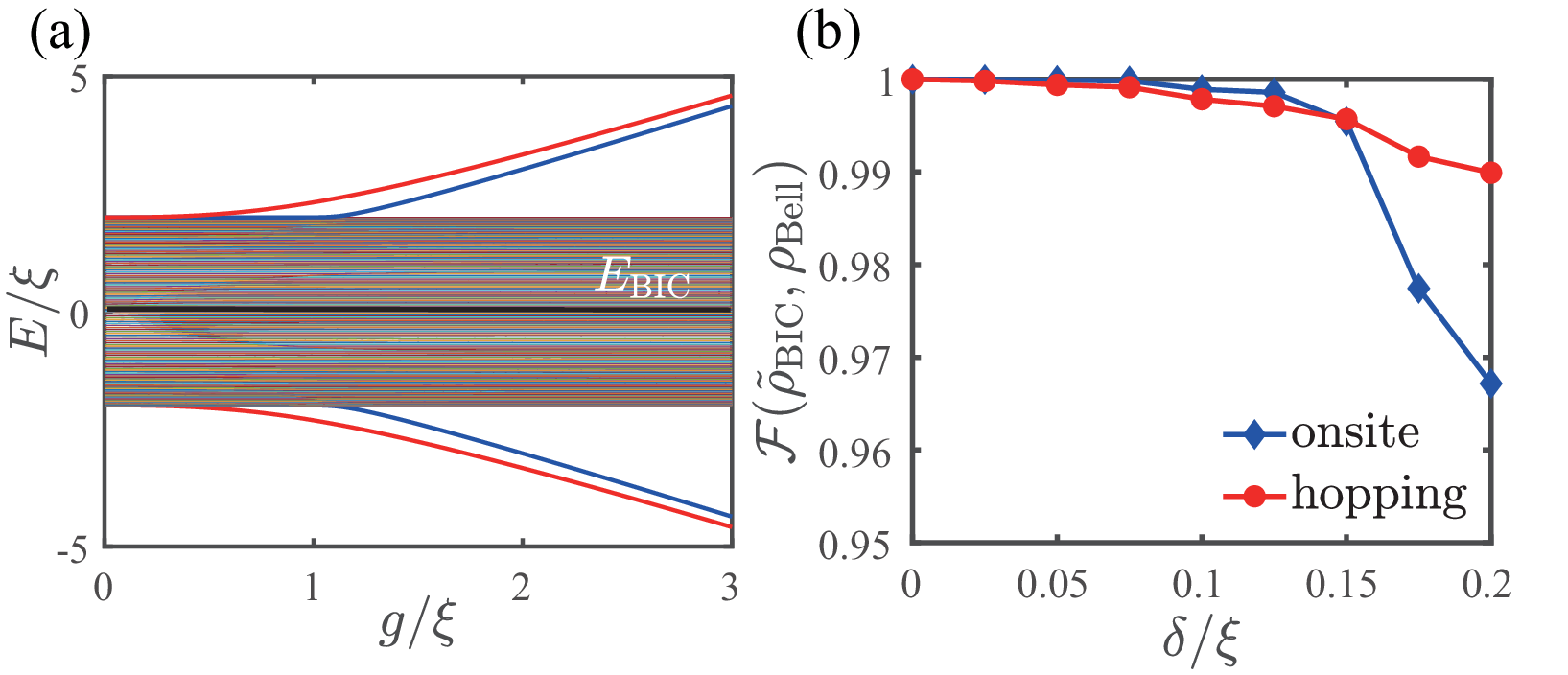}
\caption{(a) Energy diagram in the single excitation with $\omega_{c}=\Omega=0$. (b) The fidelity between $\tilde \rho_{\rm BIC}$ and $\rho_{\rm Bell}$ for onsite frequency (blue) and hopping strength (red) disorder as a function of $\delta$. The parameters are set as $\omega_{c}=\Omega$, $g=0.5\xi$, $\Delta x=2$, and $n=8$.}
\label{twoatom}
\end{figure}

Now, we can Fourier transform $a_{x}=\sum_{k}e^{-ikx}a_{k}/\sqrt{N_{c}}$ with  $N_{c}\rightarrow\infty$ being the number of the resonators, so that Eq.~(\ref{Hwaveguide}) becomes $H_{c}=\sum_{k}\omega_{k}a_{k}^{\dagger}a_{k}$, with the dispersion relation of the waveguide satisfying $\omega_{k}=\omega_{c}-2\xi\cos k$. In Fourier space, the interaction Hamiltonian reads
\begin{equation}
H_I=\sum_{k}\sum_{i=1,2}(g_{ik}\sigma_{i}^{+}a_{k}+\rm{H.c.}).
\end{equation}
Here, $g_{ik}=g(e^{-ikx_{i}}+e^{-ik(x_{i}+n)})/\sqrt{N_{c}}$ is the coupling strength, which depends on the coupling sites.
When the giant atom is resonant with the bare resonator of the waveguide, that is $\Omega=\omega_{c}$, we plot the energy spectrum of the entire system in the single excitation subspace as a function of the coupling strength $g$ as shown in Fig.~\ref{twoatom}(a). Here, the system is characterized by a continual energy band centered at $\omega_{c}$ with a total width of $4\xi$. Except for the continual band, the interaction between the giant atoms and the waveguide breaks the translational symmetry of the waveguide, leading to bound states outside of the continuum (BOC). They locate above and below the continual band, and are indicated by red and blue curves in Fig.~\ref{twoatom}(a). The BOC has been identified in various systems with structured environments, such as photonic crystals and atom-waveguide coupled systems~\cite{SJ1990,LZ2013,WZ2020}. Furthermore, the giant atoms also give rise to the BIC, which are situated within the continuous band, as represented by the black line in Fig.~\ref{twoatom}(a). When the giant atom is coupled to the 2D coupled resonator array, people also find the BIC, which bound the photon inside the atomic regime~\cite{ER2024}.

\subsection{Bound state in the continuum}

To study the BIC in detail,  we numerical obtain the single excitation eigenstate $|E_{\rm BIC}\rangle$ of the system. We find that the photon is located inside the atomic regime in the waveguide. Tracing out the bosonic modes in the waveguide, we can obtain the reduced density matrix of giant atom as $\rho_{\rm BIC}={\rm Tr}_{c}(|E_{\rm BIC}\rangle\langle E_{\rm BIC}|)$. We find that $\rho_{\rm BIC}$ is exactly the maximum entangled state, $\rho_{\rm BIC}=\rho_{\rm Bell}=|\rm Bell\rangle\langle \rm Bell|$, with $| {\rm Bell}\rangle=(|ge\rangle+|eg\rangle)/\sqrt{2}$.
This finding naturally inspires us to generate the Bell state with the BIC. However, one of the key challenges is to ensure that the BIC remains robust against disorder in the waveguide, thereby facilitating the implementation of Bell state generation in the real unideal experiment. To this end, we investigate the fidelity~\cite{NielsenBOOK}
\begin{equation}
\mathcal{F}(\tilde \rho_{\rm BIC},\rho_{\rm Bell})={\rm Tr}\sqrt{\sqrt{\tilde \rho_{\rm BIC}}\rho_{\rm Bell}\sqrt{\tilde \rho_{\rm BIC}}}.
\end{equation}
Here, $\tilde \rho_{\rm BIC}$ is the reduced density matrix of the giant atom for the BIC when the waveguide is subject to the disorder.

Concretely, we consider two kinds of disorders. One is the onsite frequency disorder while the other one is hopping strength disorder. For both cases, we model the disorder using a random distribution with standard Gaussian distribution, characterized by a full width at half-maximum of $\delta$. In Fig.~\ref{twoatom}(b), we illustrate the numerical result of the fidelity for averaging over $50$ times. The results indicate that the BIC exhibits greater robustness against hopping disorder.  Fortunately, the fidelity can be maintained above $0.96$ for both types of disorder, even when the disorder reaches  as strong as $\delta/\xi=0.2$,
which is the same order of the atom-waveguide coupling strength. Therefore, preparing the Bell state is indeed feasible. Before proposing the detailed preparation scheme, we first derive the master equation by considering the waveguide as the environment, to demonstrate the stability of the BIC.

\subsection{Master equation}

Within the Markovian approximation, the dynamics of the coupled resonator waveguide can be integrated and treated as a reservoir for the giant atoms, the formal master equation of the system in the interaction picture reads~\cite{HB}
\begin{equation}
\frac{d}{dt}{\rho}(t)=-\int_{0}^{\infty}d\tau{\rm Tr_{c}}\{[H_{I}(t),[H_{I}(t-\tau),\rho_{c}\otimes\rho(t)]]\}.
\end{equation}
Using the Wigner-Weisskopf approximation at each single coupling
point, a master equation for the density operator $\rho$ of the giant atoms can derived as~\cite{MW2024}
\begin{eqnarray}
\dot{\rho}&=&-i[H_{a},\rho]\nonumber\\
&&+\sum_{i,j}^{2}(A_{ij}+A_{ij}^{*})\sigma_{i}^{-}\rho\sigma_{j}^{+}-A_{ij}\sigma_{i}^{-}\sigma_{j}^{+}\rho-A_{ij}^{*}\rho\sigma_{i}^{-}\sigma_{j}^{+},\nonumber\\
\label{masterequation}
\end{eqnarray}
where
\begin{equation}
A_{ij}=\frac{1}{2\xi}(2e^{i\frac{\pi}{2}|x_{i}-x_{j}|}
+e^{i\frac{\pi}{2}|x_{i}+n-x_{j}|}+e^{i\frac{\pi}{2}|x_{i}-x_{j}-n|}).
\label{Aij}
\end{equation}
We rewrite the master equation in the form of $\dot{\rho}=-i[H_{\rm{eff}},\rho]+\sum_{i,j}(A_{ij}+A_{ij}^{*})\sigma_{i}^{-}\rho\sigma_{j}^{+}$, where $H_{\rm eff}$ is the effective non-Hermitian Hamiltonian
\begin{eqnarray}
H_{\rm{eff}}&=&H_{a}+\sum_{i=1}^{2}(J_{i}-i\Gamma_{i})\sigma_{i}^{+}
\sigma_{i}^{-}\nonumber \\&&+(J_{12}-i\Gamma_{12})(\sigma_{1}^{+}\sigma_{2}^{-}+\sigma_{1}^{-}\sigma_{2}^{+}).
\label{Heff}
\end{eqnarray}
Here the second term in Eq.~(\ref{Heff}) presents both the local dissipations of the giant atom with decay rate $\Gamma_{i}={\rm Re}(A_{ii})$ and the Lamb shift with the strength $J_{i}={\rm Im}(A_{ii})$. For off-diagonal term in the non-Hermitian Hamiltonian, $\Gamma_{12}={\rm Re}(A_{12})$ is the collective decay rates of the giant atoms and $J_{12}={\rm Im}(A_{12})$ is the exchanging interaction strength. For the giant atom, the coupling strength $J_{12}$ and the decay rate $\Gamma_{12}$ depend on the relative phase between the coupling sites of the giant atom, and it is not simply the sum of them at each coupling sites of that atom.

\begin{figure}
\centering
\includegraphics[width=0.85\columnwidth]{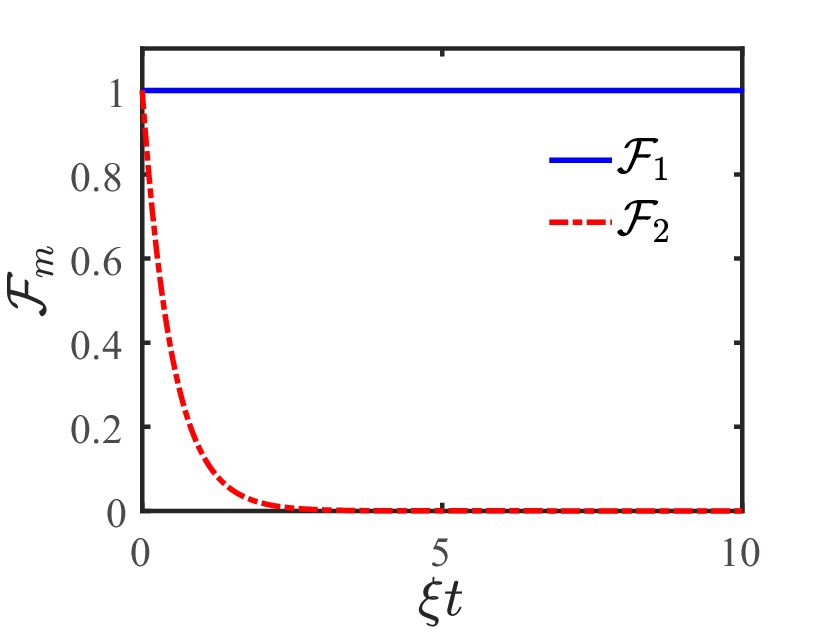}
\caption{Time evolution of the fidelity $\mathcal{F}_{m}$ for the system initially prepare in the state $\rho_{\rm BIC}$ (blue solid) and $\rho_{S}$ (red dashed). The parameter set as $\Delta x=2, n=8, \omega_{c}=\Omega$, and $g=0.5\xi$.}
\label{BICFrho}
\end{figure}

In previous work~\cite{XZ2023,XZ2024}, it is revealed that when a single giant atom is decoupled from the waveguide by the interference effect, a BIC is formed.
For the current two giant atom setup, we can still find a single BIC in the setup with $\Delta x=2$ and $n=8$.  To characterize it, we study the stability of the BIC and other scattering states. By numerically solving the master equation in Eq.~(\ref{masterequation}), we plot the time evolution of the fidelity $\mathcal{F}_m={\rm Tr}\sqrt{\sqrt{\rho_m}\rho\sqrt{\rho_{m}}}$, in which
$m=1$ and $m=2$ respectively represents that the giant atom system is initially prepared in the state
$\rho_{\rm BIC}=\rho_{\rm Bell}$ and $\rho_S={\rm Tr}_c(|S\rangle\langle S|)$ [$|S\rangle$ is an arbitrary scattering state in the continuum in Fig.~\ref{twoatom}(a)] in Fig.~\ref{BICFrho}.
 As shown by the blue solid curve, the fidelity is independent of time. This indicates that when the initial state is set as $\rho_{\rm BIC}$, it does not decay into the environment and the fidelity stabilizes at $\mathcal{F}_{1}=1$, independent of the evolution time. However, when the initial state is set as $\rho_{S}$, the fidelity decrease exponentially from $1$ to $0$ during the time evolution.

\section{Bell state generation}

The above results based on the master equation imply that the initial Bell state is immune to the dissipation induced by the coupling to the waveguide, which serves as the environment.  In other words, we can drive one of the giant atom, to prepare the two giant atom system into the Bell state. Then, we turn off the driving, the Bell state can be always alive with the assistance of the BIC. In such scheme, the driving Hamiltonian is expressed as

\begin{equation}
H_{d}=\Theta(t_{0}-t)\eta(\sigma_{1}^{+}e^{-i\omega_{d}t}
+\sigma_{1}^{-}e^{i\omega_{d}t})
\end{equation}
with $\eta$ being the Rabi frequency and $\omega_{d}$ is the frequency of the driving field. $\Theta(t_{0}-t)$ is the Heaviside function and then $t_{0}$ is the duration of driving to the first giant atom. In superconducting qubits system, the driving can be realized with independent circuits applied to the giant atoms~\cite{RB2014,PK2019}. Now the dynamics of the two giant atoms is still governed by the master equation in Eq.~(\ref{masterequation}), but $H_{a}$ is replaced by $H_{a}^{(d)}=H_{a}+H_{d}$. For a weak driving field,  the Lindblad form of Eq.~(\ref{masterequation}) dose not change.

\begin{figure}
\centering
\includegraphics[width=1\columnwidth]{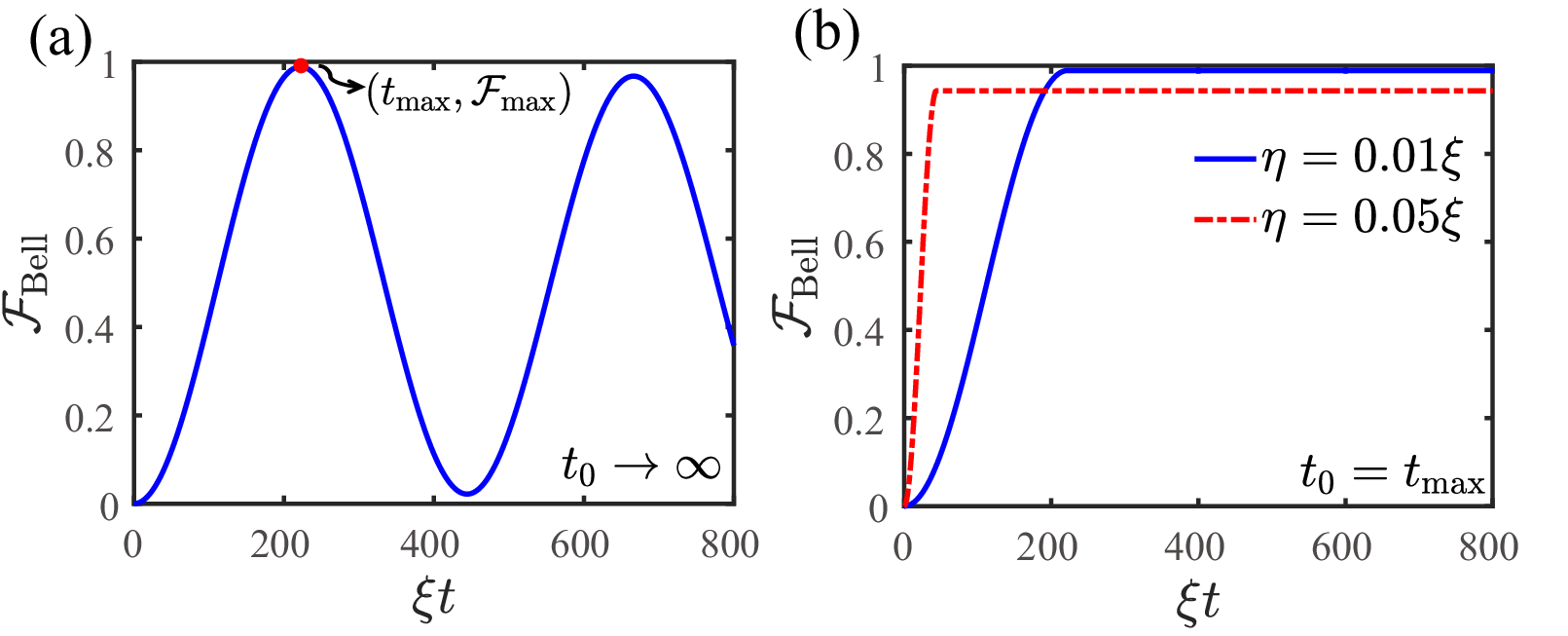}
\caption{(a) Time evolution of the fidelity $\mathcal{F}_{\rm Bell}$ with the driving strength of $\eta=0.01\xi$ and $t_{0}\rightarrow\infty$. The red dot denotes the maximum value $\mathcal{F}_{\rm max}$ of the fidelity $\mathcal{F}_{\rm Bell}$ at optimal driving duration $t_{\rm max}$. (b) Time evolution of the fidelity $\mathcal{F}_{\rm Bell}$ with the different driving strength when the duration $t_{0}=t_{\rm max}$.
The other parameters are set as $\omega_{c}=\Omega=\omega_d$ and $g=0.5\xi$.}
\label{twoatomBell}
\end{figure}

At the initial time \( t = 0 \), we prepare the two giant atoms in the ground state \( |gg\rangle = |g\rangle_1 |g\rangle_2 \). By numerically solving the master equation, we plot the time evolution of the fidelity
\begin{equation}
\mathcal{F}_{\rm Bell} = {\rm Tr}  \sqrt{\sqrt{\rho_{\rm Bell}} \, \rho \, \sqrt{\rho_{\rm Bell}}}.
\end{equation}
For long-time driving $( t_0 \rightarrow \infty )$, we observe periodic oscillations of the fidelity, as depicted in Fig.~\ref{twoatomBell}(a). The first maximum value, $ \mathcal{F}_{\rm max} \approx 1 $, occurs at the time $\xi t_{\rm max} = 223 $. This implies that the system reaches the Bell state at this moment.

This process can be intuitively decomposed as the following two steps. In the first step, the classical field drives the two atom from state $|gg\rangle$ into $|eg\rangle$. In the second step, governed by the effective Hamiltonian in Eq.~(\ref{Heff}), the interaction induces an excitation exchange between the two atoms, leading to the dressed entangled Bell states $|\Psi_{\pm}\rangle=(|eg\rangle\pm|ge\rangle)/\sqrt{2}$. The symmetric Bell state $|\Psi_+\rangle$ coincides the BIC state of the atom-waveguide coupled state and is maintained during subsequent time evolution. Meanwhile, the anti-symmetry Bell state $|\Psi_-\rangle$ will further decay back to the ground state and is subsequently re-driven. These two steps repeat multiple times, leading to the periodical oscillation in Fig.~\ref{twoatomBell}(a).

Due to the aforementioned circle, it is impossible to generate the steady Bell state with a continual driving field as $t_0\rightarrow\infty$.  To address this issue, we turn off the driving field at the moment of $t_{\rm max}$ and allow the system to evolve governed by the master equation in Eq.~(\ref{masterequation}). With the protection of the BIC, we can achieve a fidelity of $\mathcal{F}_{\rm Bell}$  as high as $100\%$ with the weak driving strength of $\eta=0.01\xi$, as shown in Fig.~\ref{twoatomBell}(b). For a stronger driving strength of $0.05\xi$, the fidelity remains above $95\%$, but the required time is significantly shortened.

\begin{figure}
\includegraphics[width=1\columnwidth]{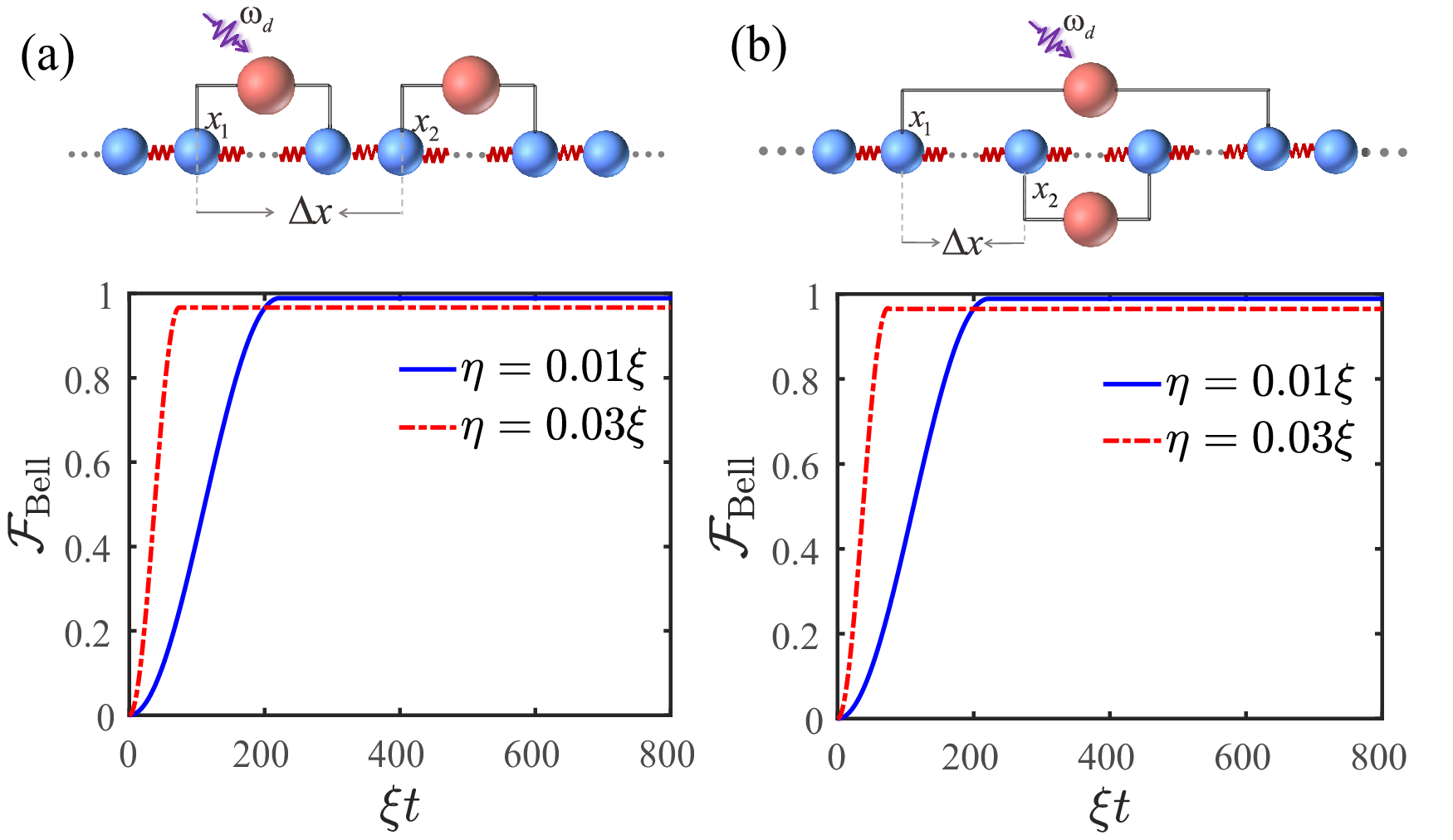}
\caption{(a) Time evolution of the fidelity $\mathcal{F}_{\rm Bell}$ with the different driving strength when the duration $t_{0}=t_{\rm max}$ in the separate configuration. The parameter set as $\Delta x=10$ and $n=8$. (b) Time evolution of the fidelity $\mathcal{F}_{\rm Bell}$ in the nested configuration. The parameter set as $\Delta x=2$, the inner atom $n=4$, and the external atom $n=8$. Other parameter set as $g=0.5\xi$.}
\label{twoarray}
\end{figure}

\section{Discussions}

In the above section, we have investigated the Bell state generation scheme in the braided configuration.  Since our scheme relies only on the BIC in the atom-waveguide coupled system, it is not limited in the braided setup. As a natural generation, we also consider the separate and nested configurations as shown in Fig.~\ref{twoarray}, which support one BIC, respectively. Using the similar strategies as before, we drive the first giant atom with an optimized during time, the corresponding Bell state preparation fidelity is demonstrated in the lower panels of Fig.~\ref{twoarray}. Again, we achieve a fidelity higher than $98\%$.

It is intuitively that the DFI cannot be over emphasized in the entangled state generation~\cite{XL2023} and such DFI can be only realized in the braided configuration~\cite{AF2018,LD2023}. Our work challenges this conventional view by the relaxing conditions for the existence of BIC. Even in our braided configuration, the effective Hamiltonian in Eq.~(\ref{Heff}) still contains the dissipation term which does not work with the DFI framework, but the BIC plays a key role to the Bell state generation. In the separate and nested configuration, we also can not find the DFI to realize a steady Bell state in the long time limit.

\begin{figure}
\includegraphics[width=1\columnwidth]{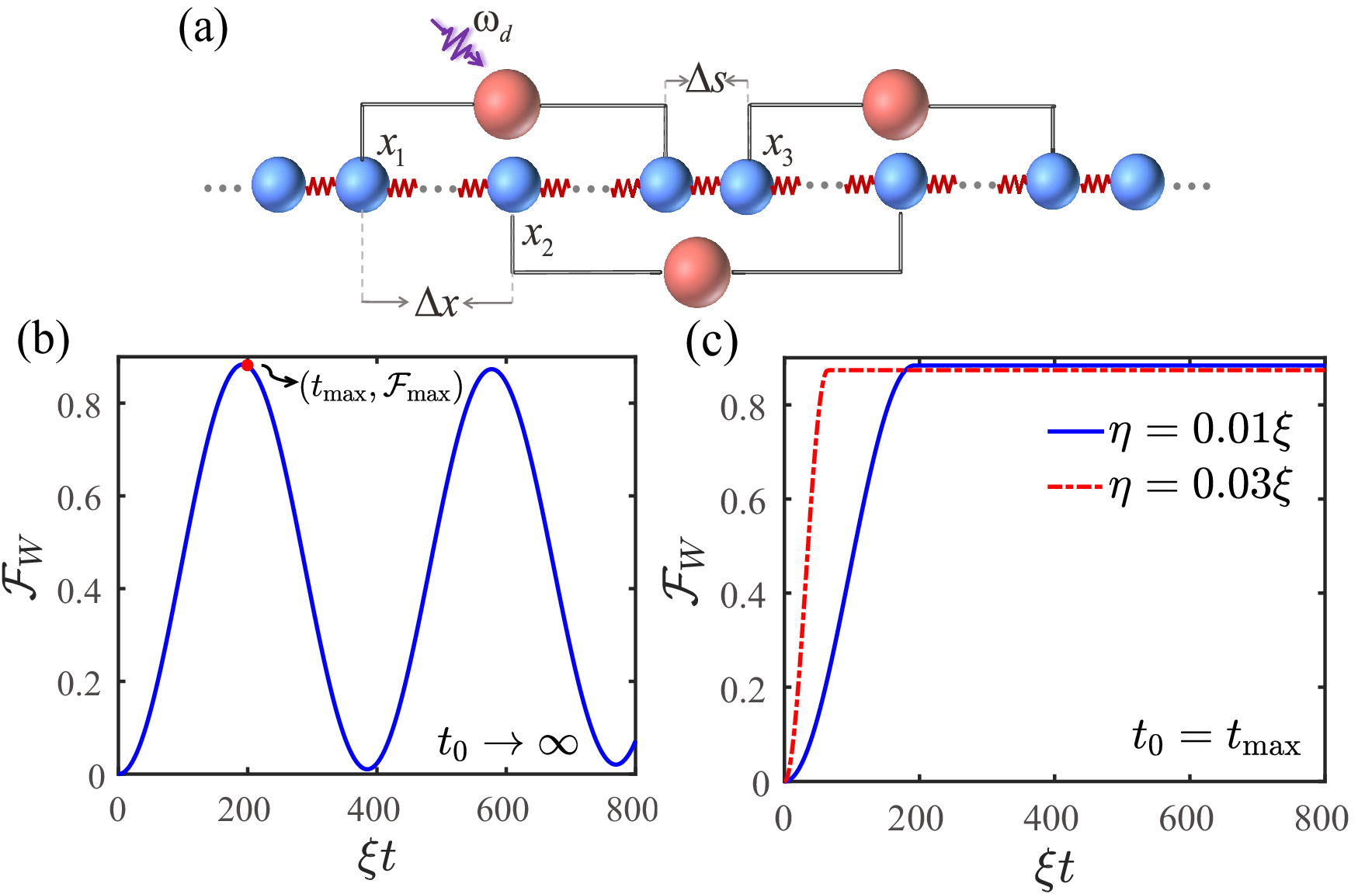}
\caption{(a) Schematic configuration for three giant atoms which couple to the one-dimensional coupled-resonator waveguide via two sites in a braided configuration. (b) Time evolution of the fidelity $\mathcal{F}_{W}$ with the driving strength $\eta=0.01\xi$ and $t_{0}\rightarrow\infty$. The red dot denotes the maximum value $\mathcal{F}_{\rm max}$ of the fidelity $\mathcal{F}_{W}$ at optimal driving duration $t_{\rm max}$. (c) Time evolution of the fidelity $\mathcal{F}_{W}$ with the different driving strength when the duration $t_{0}=t_{\rm max}$. These results are obtained in the configuration of $\Delta x=2$ and $\Delta s=2$ and the giant atom size is $n=8$. The other parameters are set as $\omega_{c}=\Omega=\omega_d$ and $g=0.5\xi$.}
\label{threeatom}
\end{figure}

Our scheme can be further extended to three giant atoms systems for the generation of three-particle W states, $|W\rangle=(|e,g,g\rangle+|g,e,g\rangle+|g,g,e\rangle)/\sqrt{3}$.
As depicted in Fig.~\ref{threeatom}(a), we consider three giant atoms coupled to the coupled resonator waveguide. The two coupled sites of each giant atom are $x_{i}$ and $x_{i}+n$ $(i=1,2,3)$. The space of neighboring giant atoms is described as $\Delta x=x_{2}-x_{1}$ and $\Delta s=x_{3}-(x_{1}+n)$. When the first giant atom is driven with Rabi frequency $\eta$, the total Hamiltonian of the system is $H=H_{a}+H_{c}+H_{I}$, where
\begin{eqnarray}
H_{a}&=&\Omega\sum_{i}^{3}\sigma_{i}^{+}\sigma_{i}^{-}+\Theta(t_{0}-t)\eta(\sigma_{1}^{+}e^{-i\omega_{d}t}+\sigma_{1}^{-}e^{i\omega_{d}t}),\nonumber\\
H_{c}&=&\omega_{c}\sum_{j}a_{j}^{\dagger}a_{j}-\xi\sum_{j}(a_{j+1}^{\dagger}a_{j}+a_{j}^{\dagger}a_{j+1}),\nonumber\\
H_{I}&=&\sum_{i}^{3}g[\sigma_{i}^{+}(a_{x_{i}}+a_{x_{i}+n})+\rm{H.c.}].
\end{eqnarray}
Under the Markovian approximation, the effective dynamics of the giant atoms is governed by the master equation
\begin{equation}
\dot{\rho}=-i[H_{a},\rho]+\mathcal{L}[\rho(t)],
\end{equation}
with the Lindblad superoperator
\begin{equation}
\mathcal{L}[\rho(t)]=\sum_{i,j}^{3}(A_{ij}+A_{ij}^{*})\sigma_{i}^{-}\rho\sigma_{j}^{+}-A_{ij}\sigma_{i}^{-}\sigma_{j}^{+}\rho-A_{ij}^{*}\rho\sigma_{i}^{-}\sigma_{j}^{+}.
\end{equation}
where $A_{ij}$ is same as Eq.~(\ref{Aij}).

At the initial time $t=0$, the system of three giant atoms is set in the ground state $|g,g,g\rangle$. With the master equation, we plot the time evolution of the fidelity $\mathcal{F}_{W}$ of the W state in Fig.~\ref{threeatom}(b) where the fidelity is defined as $\mathcal{F}_{W}={\rm Tr}\sqrt{\sqrt{\rho_{W}}\rho\sqrt{\rho_{W}}}$. Under the driving field on the first giant atom, the fidelity $\mathcal{F}_{W}$ increases monotomically to its maximum value at the optimal time $t_{\rm max}$, after which the driving field is turned off, and the W state fidelity approach $90\%$ in the long time limit, and is
nearly independent of the driving strength as shown in Fig.~\ref{threeatom}(c). Also, the scheme can be employed in the
separate and nested configuration.

\section{Conclusion}

In conclusion, we have proposed an effective scheme for generating the maximally entangled state in the giant atom waveguide system via BIC, which is robust to disorder in the waveguide. Using an external classical field, we first drive the system into the BIC, where the atomic part corresponds to an entangled state. Once the driving field is turned off, the BIC decouples from the other states of the atom-waveguide coupled system, resulting in an infinite lifetime for the atomic maximally entangled state. For the two giant atom setup, we go beyond the framework of the DFI~\cite{AF2018,LD2023} to generate a Bell state in braided, separated, and nested configurations of the giant atoms, with the fidelity exceeding $98\%$. For the three giant atom setup, we can realize the generation of a W state with a fidelity approaching $90\%$.

The giant atom can be implemented using superconducting transmon qubits~\cite{BK2020,GA2019,ZW2022}, and the coupled resonator waveguides have also been fabricated with tens of site in superconducting circuits. By expanding the capacitively coupled lumped element, nearest-neighbor hopping strengths in the range of $100$--$200$ MHz have been achieved~\cite{PR2017,XYZ2023}. Consequently, the results shown in Figs.~\ref{twoatomBell}, \ref{twoarray}, and \ref{threeatom} can be realized, even with bare qubit decay times of $T_{1} = 10\,{\rm \mu s}$~\cite{MK2020}. Our study extends beyond the DFI mechanism, providing a useful approach for constructing quantum networks and enabling quantum information processing.

\begin{acknowledgments}
Z.W. is supported by the Science and Technology Development Project of Jilin
Province (Grants No. 20230101357JC and 20220502002GH), National Science Foundation of China (Grant No. 12375010), and the Innovation Program for Quantum Science and Technology (No. 2023ZD0300700).
\end{acknowledgments}

{\bf Note} During the preparation of this work,
we find an investigation to generate the entangled states in the waveguide QED setup with the assistance of BIC via single photon scattering~\cite{GM2025}.

\end{document}